%% file: main.tex
\documentclass[conference]{IEEEtran}
\IEEEoverridecommandlockouts
\usepackage{cite}
\usepackage{amsmath,amssymb,amsfonts}

\usepackage{graphicx}
\usepackage{textcomp}
\usepackage{xcolor}
\def\BibTeX{{\rm B\kern-.05em{\sc i\kern-.025em b}\kern-.08em
    T\kern-.1667em\lower.7ex\hbox{E}\kern-.125emX}}

\usepackage{todonotes}
\usepackage{url}
\usepackage{multirow}
\usepackage{tabularray}
\usepackage{booktabs}

\usepackage{algorithm}
\usepackage{algpseudocode}

\usepackage{algpseudocode}

\makeatletter
\newcommand\fs@betterruled{%
  \def\@fs@cfont{\bfseries}\let\@fs@capt\floatc@ruled
  \def\@fs@pre{\vspace*{4pt}\hrule height.8pt depth0pt \kern3pt}%
  \def\@fs@post{\kern2pt\hrule\relax}%
  \def\@fs@mid{\kern2pt\hrule\kern2pt}%
  \let\@fs@iftopcapt\iftrue}
\floatstyle{betterruled}
\restylefloat{algorithm}
\makeatother

\usepackage{amsmath}
\usepackage{multirow}
\usepackage{subcaption}
\usepackage{comment}
\usepackage{graphicx}
\usepackage{multicol}  
\usepackage{listings}

\title{PSMOA: Policy Support Multi-Objective Optimization Algorithm for Decentralized Data Replication}
\author{
\IEEEauthorblockN{Xi Wang}
\IEEEauthorblockA{\textit{Department of Computer Science} \\
\textit{Tennessee Technological University}\\
Cookeville, USA \\
https://orcid.org/0009-0000-5169-4784}
\and
\IEEEauthorblockN{Susmit Shannigrahi}
\IEEEauthorblockA{\textit{Department of Computer Science} \\
\textit{Tennessee Technological University}\\
Cookeville, USA \\
https://orcid.org/0000-0001-6213-7473}
}
\IEEEaftertitletext{\vspace{-2\baselineskip}}

\begin{document}

\IEEEoverridecommandlockouts
%\IEEEpubid{\makebox[\columnwidth]{ISBN 978-3-903176-57-7\copyright 2023 IFIP \hfill} \hspace{\columnsep}\makebox[\columnwidth]{ }}
\maketitle

\input{abstract.tex}


\input{0-Intro+background}
%\input{0-Introduction}
\input{1-Background}

\input{2-GNSGA2}
\input{5-Experiments-and-Simulation}

\input{6-Conclusions}

%\section*{Acknowledgments}
%This work has been supported by National Science Foundation Awards OAC-2019163, OAC-2126148, DGE-2043324, and OAC-2019012.

\section{Acknowledgments}
This material is based upon work supported by the National Science Foundation under Grant Numbers 2019163 and 2126148.

\bibliographystyle{unsrt}
\bibliography{references.bib}

%\tableofcontents

\end{document}

%% file: abstract.tex
\begin{abstract}

%Efficient data replication is crucial for decentralized storage federations used in data-intensive scientific collaborations, where each organization has distinct operational policies. 

Efficient data replication in decentralized storage systems must account for diverse policies, especially in multi-organizational, data-intensive environments. This work proposes PSMOA, a novel Policy Support Multi-objective Optimization Algorithm for decentralized data replication that dynamically adapts to varying organizational requirements 
%. PSMOA integrates NSGA-III with Entropy Weighted TOPSIS to optimize replication 
such as minimization or maximization of replication time, storage cost, replication based on content popularity, and load balancing while respecting policy constraints. %Our simulations demonstrate PSMOA's superior performance, with load balancing %maintaining 104-107\% 
%performance improving by 4-7\% relative to baseline.
%, while other metrics show stable performance between 98-103\%. 
PSMOA outperforms NSGA-II and NSGA-III in both Generational Distance (20.29 vs 148.74 and 67.74) and Inverted Generational Distance (0.78 vs 3.76 and 5.61), indicating better convergence and solution distribution. These results validate PSMOA's novelty in optimizing data replication in multi-organizational environments.

\end{abstract}

%% file: 0-Intro+background.tex
\section{Introduction}
%In the era of big data, managing and processing vast amounts of information in large-scale distributed systems has become increasingly challenging. The Worldwide LHC Computing Grid (WLCG) supports the Large Hadron Collider (LHC) experiments at CERN \cite{cern}, exemplifies these challenges, where effective data replication strategies are essential for system performance. 

%In the era of big data, 
Data replication is crucial for availability, reliability, and performance in large-scale, multi-organizational distributed systems\cite{liu2021designing, wu2022n}. However, different nodes in such systems may operate under various constraints and policies, making policy-aware replication a complex challenge\cite{wang2023gnsga}. The Worldwide LHC Computing Grid (WLCG), which supports the Large Hadron Collider (LHC) experiments at CERN \cite{wu2022n}, exemplifies these challenges. It distributes vast amounts of data across globally dispersed nodes where each node may have its own distinct policies and constraints. Limited research has been dedicated to addressing policy constraints in such decentralized, multi-organizational data replication.

\begin{figure}[h]
\centering
\includegraphics[width=\columnwidth]{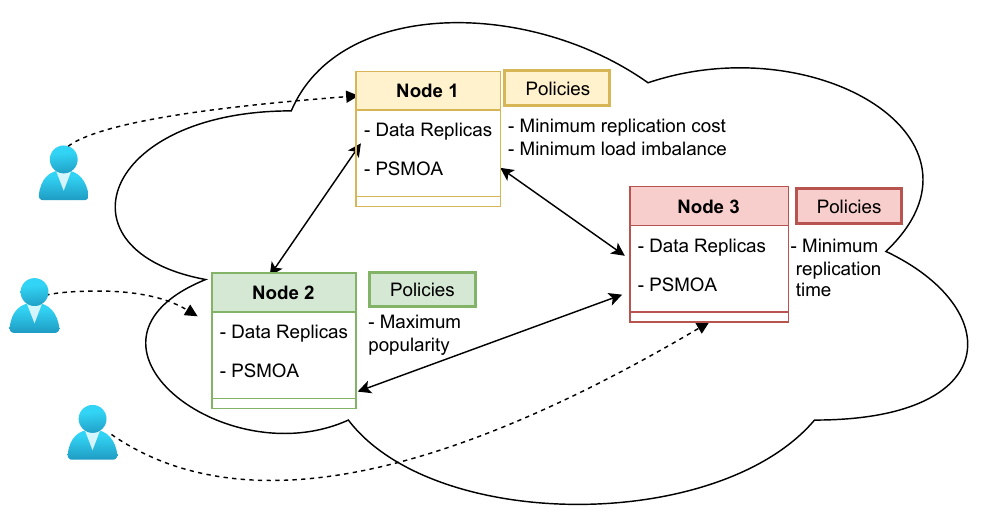}
\caption{Decentralized data replication system model.}
\label{fig:model}
\vspace{-1em}
\end{figure}

Fig. \ref{fig:model} shows a model of a decentralized data replication system in which multiple nodes collaborate to replicate data. %by replicating them across organizations 
while maintaining local policies. In this example, Node 1's policy is to minimize replication cost and storage imbalance, Node 2 aims to maximize local replication of popular datasets, and Node 3 aims to reduce replication time. Traditional replication decisions typically optimize for a single goal and may not work well for this example. 
%Further, goals and constraints may change while the nodes are replicating. 

This work proposes the Policy Support Multi-Objective Optimization Algorithm (PSMOA), a novel approach that integrates NSGA-III's capabilities in handling many-objective optimization problems with dynamic policy support.
%through the Entropy Weighted TOPSIS method. 
The novelty of this work is three-fold. First, we introduce a flexible framework for expressing and dynamically adjusting organizational policies. Second, our algorithm effectively handles the inherent trade-offs between competing objectives while maintaining solution diversity. Third, by enabling independent node operation while preserving coordinated optimization, PSMOA avoids the limitations of centralized approaches. Our simulations show 
%PSMOA's superior performance, with load balancing performance improving by 4-7\% relative to baseline. 
the algorithm consistently maintains performance metrics above baseline (104-107\% for load balancing, 98-103\% for other objectives) while adapting to varying policy requirements. Performance comparisons with existing approaches show that PSMOA achieves better convergence (GD: 20.29) and solution distribution (IGD: 0.78).
%compared to existing approaches.

%The algorithm also outperforms existing approaches in both convergence and solution distribution metrics.

%% file: 1-Background.tex
\section{Related work}

Multi-objective optimization problems in decentralized data replication present significant challenges in properly weighing different objectives. Equally weighting objectives that are inherently connected may not properly reflect their interdependencies \cite{verma2021comprehensive}.
This challenge is particularly evident in data replication scenarios where objectives like cost, time, and popularity often have complex relationships.
%The emergence of advanced multi-objective optimization algorithms has opened new possibilities. 
While NSGA-III \cite{deb2013evolutionary} effectively handles many objectives and maintains well-distributed Pareto fronts, its application to data replication requires careful consideration of objective weights and policy integration.
Recent work by Blondin and Hale \cite{blondin2021decentralized} and Shukla and Pandey \cite{shukla2024motors} addressed multi-objective optimization in decentralized systems, 
but both approaches lack mechanisms for dynamic policy adaptation in real-world scenarios. 
Our previous work \cite{wang2023gnsga} uses a favor-based approach for multi-objective optimization in data replication decisions. Such approaches can be used to build real-world decentralized storage systems such as Hydra \cite{presley2024hydra}, a peer-to-peer storage federation that enables decentralized and reliable data publication capabilities, and NDISE \cite{wu2022n, 10820568} that creates a mechanism for high-speed data transfer mechanisms between LHC sites.
%but with different limitations. Blondin and Hale's approach lacks practical implementation details and remains largely theoretical. Similarly, MOTORS framework's evaluation was limited to specific benchmark workflows and may not generalize well to dynamic real-world scenarios where workload patterns and resource availability constantly change.
Previous studies have applied NSGA-II variants to solve multi-objective combinatorial optimization problems\cite{verma2021comprehensive}. However, the integration of policy support with many-objective optimization for data replication remains unexplored. %Additionally, existing approaches often struggle with dynamic policy adaptation and objective weight assignment in decentralized environments.
To address these limitations, we incorporate user policies through the Entropy Weighted TOPSIS method \cite{chen2021effects}, using information content-based weights to guide the search process, toward optimal data replication solutions in decentralized environments.

%% file: 2-GNSGA2.tex
\section{PSMOA}

%\todo[inline]{Can you give an example of couple of iterations that shows how the weights are changing? We can discuss this tomorrow morning}
The core idea for the PSMOA Algorithm is to integrate user preferences and policies into the NSGA-III framework, making it adaptable to various decentralized and federated systems. 
The pseudocode in Algorithm \ref{algo:algo1} begins by initializing user preference factors $\alpha$ and running Entropy Weighted TOPSIS to obtain initial weights $w$. It then checks if a user has specified a local policy. If no policy is specified, the weights remain unchanged ($w' = w$). If a single-objective policy is specified, the algorithm applies single-objective optimization and returns the best solution, ending the process. For a multi-objective policy, the algorithm updates $\alpha$ based on the policy and adjusts weights $w$ using the updated $\alpha$.
After policy handling, the algorithm generates reference points $Z$ using the adjusted weights $w'$. The core optimization process is then executed using NSGA-III. 
The optimization loop follows the NSGA-III structure, creating offspring, performing non-dominated sorting, and selecting the next generation based on reference points and niche counts. 
\begin{algorithm}[!h]
\caption{PSMOA Algorithm}
\begin{algorithmic}[1]
\State Initialize user preference factors $\alpha = \{\alpha_1, \alpha_2, \ldots, \alpha_m\}$ to 1
\State Run Entropy Weighted TOPSIS to get initial weights $w = \{w_1, w_2, \ldots, w_m\}$
\If{user specifies a policy}
    \If{policy is single-objective}
        \State Apply single-objective optimization
        \State \Return best solution
    \ElsIf{policy is multi-objective}
        \State Update user preference factors $\alpha$ based on policy
        \State Adjust weights $w$ using $\alpha$:
        \State $w'_j = \frac{\alpha_j w_j}{\sum_{j=1}^{m} \alpha_j w_j}, \quad j \in [1, m]$
    \EndIf
\Else
    \State $w' \gets w$
\EndIf
\State Generate reference points $Z$ using adjusted weights $w'$
\State Initialize population $P_0$ of size $N$
\State $t \gets 0$
\While{termination criteria not met}
    \State $Q_t \gets$ Create offspring population from $P_t$
    \State $R_t \gets P_t \cup Q_t$
    \State $F = (F_1, F_2, \ldots) \gets$ Non-dominated-sort($R_t$)
    \State Normalize objectives and reference points
    \State Associate each member of $F$ with a reference point
    \State Compute niche count of reference points
    \State $P_{t+1} \gets$ Select $N$ population members based on niche count and association
    \If{policy changes}
        \State Update user preference factors $\alpha$ based on new policy
        \State Adjust weights $w'$ using updated $\alpha$
        \State Update reference points $Z$ using adjusted weights $w'$
    \EndIf
    \State $t \gets t + 1$
\EndWhile
\State \Return $P_t$ as the optimal set of solutions
\end{algorithmic}
\label{algo:algo1}
\end{algorithm} 

Crucially, the algorithm checks for policy changes at each iteration, adjusting weights and updating reference points if necessary. This approach allows for dynamic adaptation to changing user preferences or system conditions.
Finally, the algorithm returns the optimal set of solutions found after the termination criteria are met. 
%This approach ensures that the algorithm remains flexible and applicable to a wide range of decentralized systems, adapting to different policies and preferences while maintaining the core strengths of multi-objective optimization.

\begin{comment}
\begin{figure}[h]
\centering
\includegraphics[width=0.8\columnwidth, trim={1cm 1cm 1cm 0}]{figs/flow1.pdf}
\caption{PSMOA Flowchart}
\label{fig:flow}
\vspace{-2em}
\end{figure}

The PSMOA algorithm flowchart in Fig. \ref{fig:flow} begins with initializing user preference factors $\alpha$. It then runs Entropy Weighted TOPSIS to obtain initial weights $w$. The process then checks if a user specifies a policy.
If no policy is specified, the weights remain unchanged ($w' = w$). If a policy is specified, the algorithm branches based on the policy type: 1. for a single-objective policy, it applies single-objective optimization and returns the best solution, ending the process. 2. for a multi-objective policy, it updates $\alpha$ based on the policy and adjusts weights $w$ using the updated $\alpha$.
After policy handling, the algorithm generates reference points $Z$ using the adjusted weights $w'$. 
The core optimization process is then executed using NSGA-III.
Finally, the algorithm returns an optimal set of solutions and ends.
\end{comment}

\subsection{Entropy Weighted TOPSIS} \label{A}
Entropy Weighted TOPSIS combines Shannon's entropy weight theory \cite{shannon1948mathematical} and the Technique for Order Preference by Similarity to Ideal Solution (TOPSIS) \cite{hwang2012multiple} approach to generate reference points for multi-objective optimization. Shannon's entropy concept provides a measure of information content and uncertainty in data \cite{hwang2012multiple}, which is used to objectively determine weights for different objectives based on their information content. TOPSIS evaluates alternatives by measuring their distances from ideal and anti-ideal solutions \cite{kahraman2008multi}.

\subsubsection{Objective Weight Assignment}
By calculating the entropy of each objective, we can quantify the amount of information it provides and assign higher weights to objectives with lower entropy, thereby prioritizing the most informative and discriminative objectives. 

The objective weight for each criterion $j$ is given by:
$w_j = \frac{d_j}{\sum_{j=1}^{n} d_j}, \quad j \in [1, n]$ 
where $d_j = 1 - e_j$ represents information divergence, with entropy $e_j$ calculated from normalized performance ratings $p_{ij}$ as:
$e_j = -\frac{1}{\ln(m)} \sum_{i=1}^{m} p_{ij} \ln(p_{ij}), \quad i \in [1, m], j \in [1, n]$, where $m$ is the number of alternatives, and $n$ is the number of objectives.

As a policy support algorithm, PSMOA allows users to express their preferences or policies by adjusting these objective weights. %For example, if a user prioritizes lower replication costs, they can increase the weight assigned to the cost objective. In the absence of user-specified policies, PSMOA defaults to the weights determined by the Entropy Weighted TOPSIS method.
To incorporate user preferences, we introduce a weight adjustment factor $\alpha_j$ for each objective $j$. The adjusted weight $w'_j$ is calculated as:
$w'_j = \frac{\alpha_j wj}{\sum_{j=1}^{n} \alpha_j w_j}, \quad j \in [1, n]$ 
where $\alpha_j \geq 0$ represents the user's preference for objective $j$. A higher value of $\alpha_j$ indicates a stronger preference for the corresponding objective. If no user preferences are specified, $\alpha_j = 1$ for all objectives, and the weights remain unchanged.
The adjusted objective weights $w'_j$ are then used to generate the reference points for the NSGA-III algorithm.

\subsubsection{Reference Point Generation}
The reference points are generated to reflect the relative importance of each objective and guide the search toward regions of the objective space that are more informative and potentially more important, based on the inherent characteristics of the problem data. The method consists of the following steps:

\paragraph{Determine the positive ideal solution $Z^+$ and negative ideal solution $Z^-$ in the objective space}
$Z^\pm = \left(
\begin{aligned}
\max(\min)\{z_{11}, z_{21}, \dots, z_{n1}\}, & \\
\max(\min)\{z_{12}, z_{22}, \dots, z_{n2}\}, & \\
\dots, & \\
\max(\min)\{z_{1m}, z_{2m}, \dots, z_{nm}\} &
\end{aligned}
\right) = (Z^\pm_1, Z^\pm_2, \dots, Z^\pm_m)$

\begin{comment}
$Z^\pm = \left(
\begin{aligned}
\max(min)\{z_{11}, z_{21}, \dots, z_{n1}\}, & \\
\max(min)\{z_{12}, z_{22}, \dots, z_{n2}\}, & \\
\dots, & \\
\max(min)\{z_{1m}, z_{2m}, \dots, z_{nm}\} &
\end{aligned}
\right) = (Z^\pm_1, Z^\pm_2, \dots, Z^\pm_m)$

\end{comment}

\paragraph{Calculate the closeness of each solution to the positive and negative ideal solutions} $D_i^\pm = \sqrt{\sum_{j=1}^m w'_j (Z_j^\pm - z_{ij})^2}$
, where $D_i^+$ represents the distance from the $i$-th solution to the ideal solution, $D_i^-$ represents the distance to the negative ideal solution, $Z_j^+$ and $Z_j^-$ are the ideal and anti-ideal values for the $j$-th objective, $z_{ij}$ is the value of the $i$-th solution for the $j$-th objective, and $w'_j$ is the weight obtained from the entropy method.

\paragraph{Compute the relative closeness coefficient $C_i$ for each solution} $C_i = \frac{D_i^-}{D_i^+ + D_i^-}$.
The larger the $C_i$ value, the closer the Pareto solution is to the positive ideal solution, indicating a better compromise among the conflicting objectives.

\paragraph{Generate reference points based on the relative closeness coefficients and entropy weights}
The reference points are distributed in the objective space, considering the importance of each objective determined by the entropy weights and the closeness of solutions to the ideal solution.

%For moderate indicators, where neither extremely high nor low values are desirable, a normalization process is applied. Let ${x_i}$ be a set of moderate indicator sequences, and $x_{ideal}$ be the ideal value. The normalization formula is $M = \max \{\|x_i - x_{ideal}\|\}, \quad \tilde{x_i} = 1 - \frac{\|x_i - x_{ideal}\|}{M}, \quad i=1,2,\dots,n$, where $x_i$ is the original value of the $i$-th sample. $x_{ideal}$ is the ideal value for the indicator. $M$ is the maximum absolute deviation from the ideal value across all samples. $\tilde{x_i}$ is the normalized value of the $i$-th sample. This normalization ensures that values closer to the ideal are transformed to be closer to 1, while those further away are closer to 0.

By incorporating user preferences, PSMOA offers a customizable approach to multi-objective optimization for data replication in decentralized systems, adapting to specific requirements and policies while leveraging the Entropy Weighted TOPSIS method for objective weight assignment.

\subsection{Policy Expression}
PSMOA provides a flexible framework for expressing policies through weight vectors and constraint specifications. Organizations express their operational requirements and preferences through a combination of objective weights and system constraints. %The policy specification process consists of weight vector and dynamic weight adjustment.
The policy preferences are primarily expressed through a weight vector $\boldsymbol{\alpha} = [\alpha_T, \alpha_C, \alpha_P, \alpha_L]$, these weights are normalized such that $\sum\alpha_i = 1$, and $\alpha_i \in [0,1]$. This ensures a balanced representation of priorities. 
%\todo[inline]{check this - doesn't look right}
%$\alpha$ = [$\alpha$time, $\alpha$cost, $\alpha$popularity, $\alpha$load]. These weights reflect the relative importance assigned to each objective based on organizational priorities.

The final weights used in the optimization process are calculated through a dynamic adjustment mechanism that combines the policy weights with entropy-based weights. Specifically, the adjusted weight $w'_j$ is calculated as $w'_j = (\alpha_j \times w_j) / \sum(\alpha_j \times w_j)$, where $w_j$ is the entropy-based weight reflecting the inherent information content of each objective. 

For instance, a research institution might specify $\boldsymbol{\alpha} = [0.4, 0.2, 0.3, 0.1]$, to emphasize replication time, while a data center might use $\boldsymbol{\alpha} = [0.2, 0.4, 0.2, 0.2]$ to prioritize cost optimization.

\subsubsection{Dynamic Policy Adaptation}

PSMOA continuously adapts policies by modifying the weight vector $\boldsymbol{\alpha}$ based on system conditions. The adaptation process adjusts individual components of the weight vector while maintaining the normalization constraint $\sum \alpha_i = 1$.

For load balancing, the system adjusts $\alpha_L$ during high-usage periods according to utilization rates: $\alpha_L = \min(\alpha_{L\_base} + \lambda \times \text{utilization\_rate}, \alpha_{\text{max}})$, where $\lambda$ represents the adaptation rate and $\text{utilization\_rate}$ reflects current system usage. 

The cost component $\alpha_C$ is adjusted as resource usage approaches predefined limits: $\alpha_C = \alpha_{C\_base} \times (1 + \beta \times \text{budget\_proximity})$
, where $\text{budget\_proximity}$ indicates how close the system is to its budget constraints. 

For data popularity considerations, PSMOA adapts $\alpha_P$ based on observed access patterns: $\alpha_P = \alpha_{P\_base} \times (1 + \gamma \times \text{access\_frequency})$
, where $\text{access\_frequency}$ is normalized against historical data, and $\gamma$ is the popularity sensitivity coefficient. 

After each adaptation step, the weights are renormalized to maintain the constraint $\sum \alpha_i = 1$: $\alpha_i' = \alpha_i / \sum_{j \in \{T,C,P,L\}} \alpha_j$.

\subsubsection{User-Imposed Constraints}

Users can impose various constraints through a policy specification interface. Resource constraints form the foundation of policy expression, allowing organizations to define storage limits ($max\_storage\_per\_node \leq storage\_capacity$), bandwidth thresholds ($replication\_rate \leq available\_bandwidth$), and cost boundaries ($monthly\_cost \leq budget\_limit$). 
Quality of service requirements are expressed through constraints on maximum replication time ($replication\_time \leq max\_allowed\_time$), minimum availability ($replicas\_count \geq min\_replicas$), and geographic distribution of data($region\_diversity \geq min\_regions$).

Users can further define conditional policies using if-then rules that reflect specific operational requirements. Critical data might require a minimum of three replicas and strict replication time limits, for example, 
(\verb|if data_type is `critical': min_replicas=3|
\verb|and max_replication_time = `1hr'|),
% \begin{lstlisting}
% if data_type is `critical':
%    min_replicas=3
%    and max_replication_time = `1hr'
% \end{lstlisting}
% elif data_size > 1TB:
%     cost_weight = 0.5
%     load_weight = 0.3
while large datasets above certain size thresholds might trigger adjusted weight distributions for cost and load considerations. These policy expressions are integrated into PSMOA's optimization process through the entropy-weighted approach discussed in Section \ref{A}.

\subsection{NSGA-III Algorithm for Node Selection} 
NSGA-III is employed to handle many-objective optimization problems, including potential applications like data replication in decentralized systems. It enhances the selection mechanism to ensure better diversity among solutions by utilizing a set of reference points, maintaining a well-distributed population by associating, and improving convergence through non-dominated sorting, elitist selection, and niching. The multi-objective optimization model for data replication in the decentralized system is formulated as follows:

\begin{equation}\label{eq:multi}
\begin{aligned}
F(X) &= [T(X), C(X), -P(X), L(X)] \\
\text{subject to:}\quad g_i(X) &\leq 0, \quad i = 1, 2, \ldots, m \\
h_j(X) &= 0, \quad j = 1, 2, \ldots, p
\end{aligned}
\end{equation}

where $X$ is the decision vector representing the replication strategy, $T(X)$ is the replication time objective, $C(X)$ is the replication cost objective, $P(X)$ is the popularity objective, $L(X)$ is the load balance objective, $g_i(X)$ are the inequality constraints, and $h_j(X)$ are the equality constraints. Equation \ref{eq:multi} represents the multi-objective optimization problem.

The measurement metrics used in our optimization approach are grounded in established research on distributed data systems. Our replication time objective T(X) builds on Rahman et al.'s network transfer model \cite{rahman2008replica}, incorporating both bandwidth and latency components that have been validated in data grid environments. 
The cost metric C(X) follows Khan et al.'s comprehensive cloud storage cost taxonomy \cite{khan2024cloud}, which establishes that total storage cost must account for both direct storage costs and related costs like data replication and network usage. 
For measuring popularity P(X), we adopt the access pattern modeling approach validated by Kroeger and Long \cite{kroeger1999case}, who demonstrated that file accesses follow strong correlative patterns that can be effectively measured and predicted. 
Our load balancing metric L(X) builds on Guo et al.'s variance-based approach \cite{guo2014improving}, which demonstrated the effectiveness of using load variance thresholds to achieve balanced resource utilization in distributed systems. 

%To solve this optimization problem effectively, we define each objective function to reflect real-world requirements.

\subsubsection{Optimization Objectives} 
\label{objs}

The optimization model addresses four critical objectives simultaneously in decentralized data replication systems:

\paragraph{Replication time objective}
In data-intensive scientific collaborations like the WLCG, minimizing replication time is crucial for maintaining data availability and research productivity. Network conditions and node capabilities significantly impact transfer speeds, making it essential to consider both bandwidth limitations and latency effects. 
The time objective $T(X)$ minimizes the total time required to replicate data objects across the selected nodes in the system, incorporating network latency through:

\begin{equation}\label{eq:time}
\min T(X) = \sum_{j \in S_i} (\frac{C_N}{B_j} + RTT_{uj})
\end{equation}

where $S_i$ is the set of selected nodes for data object $i$, $C_N$ is the size of the data object, $B_j$ is the bandwidth of node $j$, and $RTT_{uj}$ is the round-trip time between the user node $u$ and node $j$. The inclusion of network latency, represented by the round-trip time (RTT), captures the real-world impact of network infrastructure on replication performance.

\paragraph{Replication cost objective}

With the exponential growth of scientific data volumes, cost optimization becomes increasingly critical for the sustainable operation of distributed storage systems. %Storage expenses and network transfer charges can significantly impact operational budgets. 
The cost objective $C(X)$ minimizes the total cost associated with the replication process %by calculating the total cost, 
which encompasses both storage and data transfer costs:

\begin{equation}\label{eq:cost}
\min C(X) = |S_i| \times (c_1 \times C_N + c_2 \times C_N)
\end{equation}

where $|S_i|$ represents the cardinality or the number of elements in the set $S_i$, $c_1$ is the storage cost coefficient and $c_2$ is the data transfer cost coefficient.
This formulation reflects the economic realities of modern data centers, where both storage capacity and network bandwidth contribute substantially to operational costs.

\paragraph{Popularity objective}
Data access patterns in scientific collaborations often show strong temporal and spatial locality, with certain datasets experiencing high demand during specific research phases. %Optimizing replica placement based on popularity improves overall system efficiency. 
The popularity objective $P(X)$ aims to maximize the total popularity score of the selected nodes for each data object:

\begin{equation}\label{eq:popularity}
\max P(X) = \sum_{j=1}^{n} X_{ij} \times p_j
\end{equation}

where $X_{ij}$ is the decision variable indicating whether node $j$ is selected for data object $i$ (1 if selected, 0 otherwise), and $p_j$ is the popularity score of node $j$. The popularity score is calculated based on the number of requests for the data object, enabling the system to adapt to dynamic access patterns.

\paragraph{Load balance objective}
In distributed systems, uneven resource utilization can lead to performance bottlenecks and reduced system efficiency. %Maintaining balanced load distribution is essential for consistent performance and resource utilization. 
The load balance objective $L(X)$ aims to minimize the variance of the load distribution among the nodes in the system:

\begin{equation}\label{eq:load}
\min L(X) = \frac{1}{n} \sum_{j=1}^{n} (l_j - \bar{l})^2
\end{equation}

where $l_j$ is the load on node $j$, $\bar{l}$ is the average load across all nodes, and $n$ is the total number of nodes in the system. This formulation ensures that no single node becomes a bottleneck, promoting consistent performance across the distributed infrastructure.

\subsubsection{Constraints}

The optimization model is subject to various constraints to ensure the feasibility and practicality of the replication strategy. The constraints are categorized as follows:

\paragraph{Storage capacity constraint}
The total size of replicas stored on each node should not exceed its storage capacity. This constraint can be expressed as $\sum_{i=1}^{m} X_{ij} \times C_N \leq N_j, \quad \forall j = 1, 2, \ldots, n$
, where $X_{ij}$ is the decision variable indicating whether node $j$ is selected for data object $i$, $C_N$ is the size of the data object, and $N_j$ is the storage capacity of node $j$.

\paragraph{Bandwidth constraint}
The total bandwidth consumed by the replication process should not exceed the available bandwidth of each node. This constraint is formulated as
$\sum_{i=1}^{m} X_{ij} \times C_N \leq B_j, \quad \forall j = 1, 2, \ldots, n$
, where $B_j$ is the available bandwidth of node $j$.

%% file: 5-Experiments-and-Simulation.tex
\section{Simulation} 

\begin{figure*}[!ht]
    \centering
    \begin{subfigure}[b]{0.329\textwidth}
        \centering
        \includegraphics[width=\textwidth]{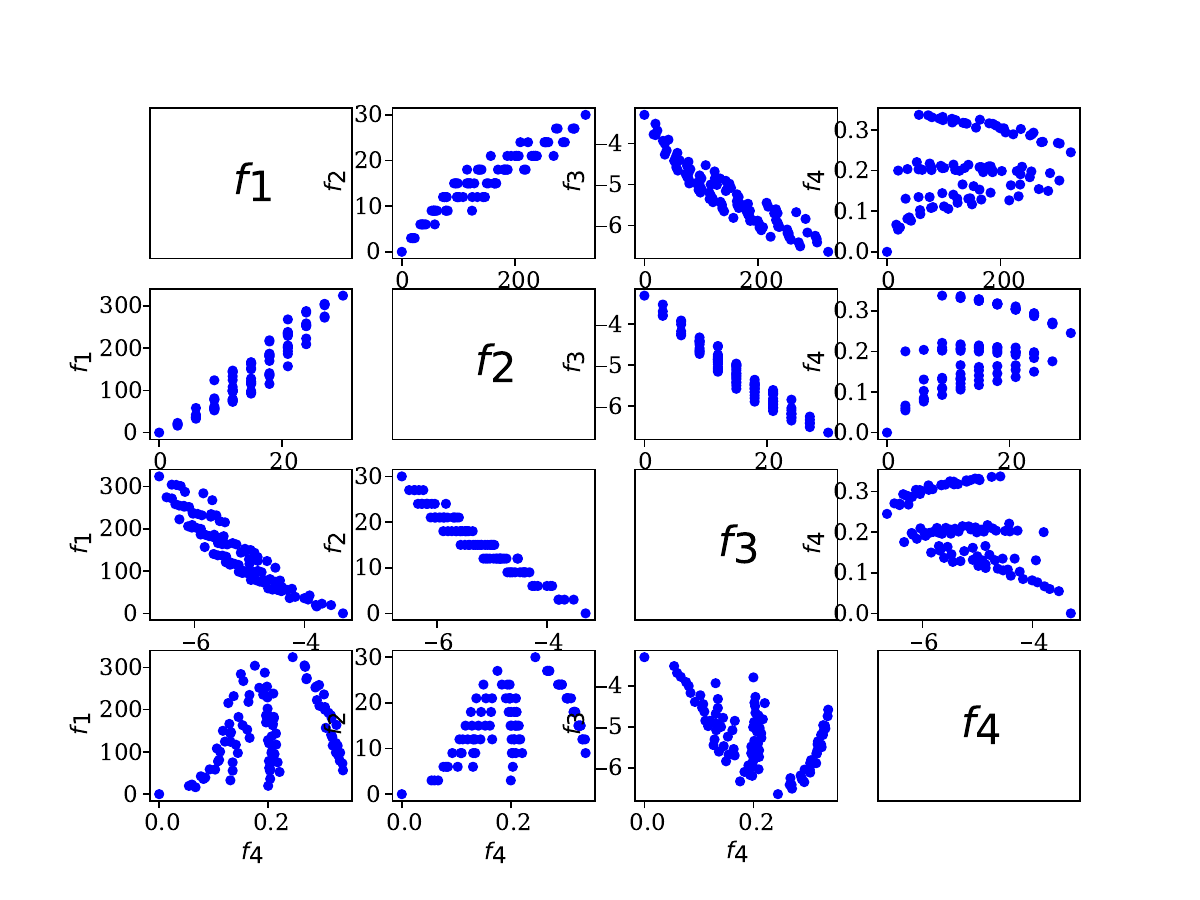}
        \caption{NSGA-II}
        \label{fig:figure1}
    \end{subfigure}
        \begin{subfigure}[b]{0.329\textwidth}
        \centering
        \includegraphics[width=\textwidth]{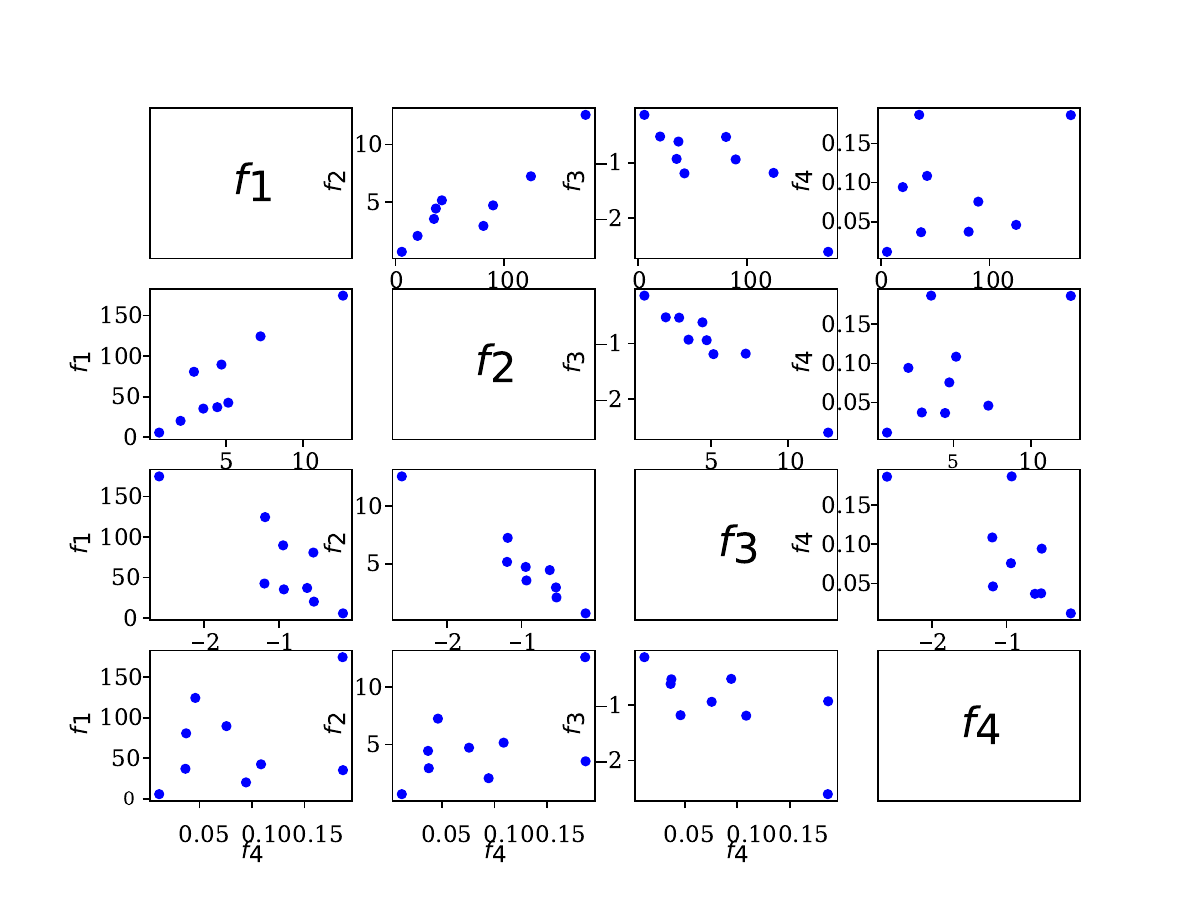}
        \caption{NSGA-III}
        \label{fig:figure2}
    \end{subfigure}
    \begin{subfigure}[b]{0.329\textwidth}
        \centering
        \includegraphics[width=\textwidth]{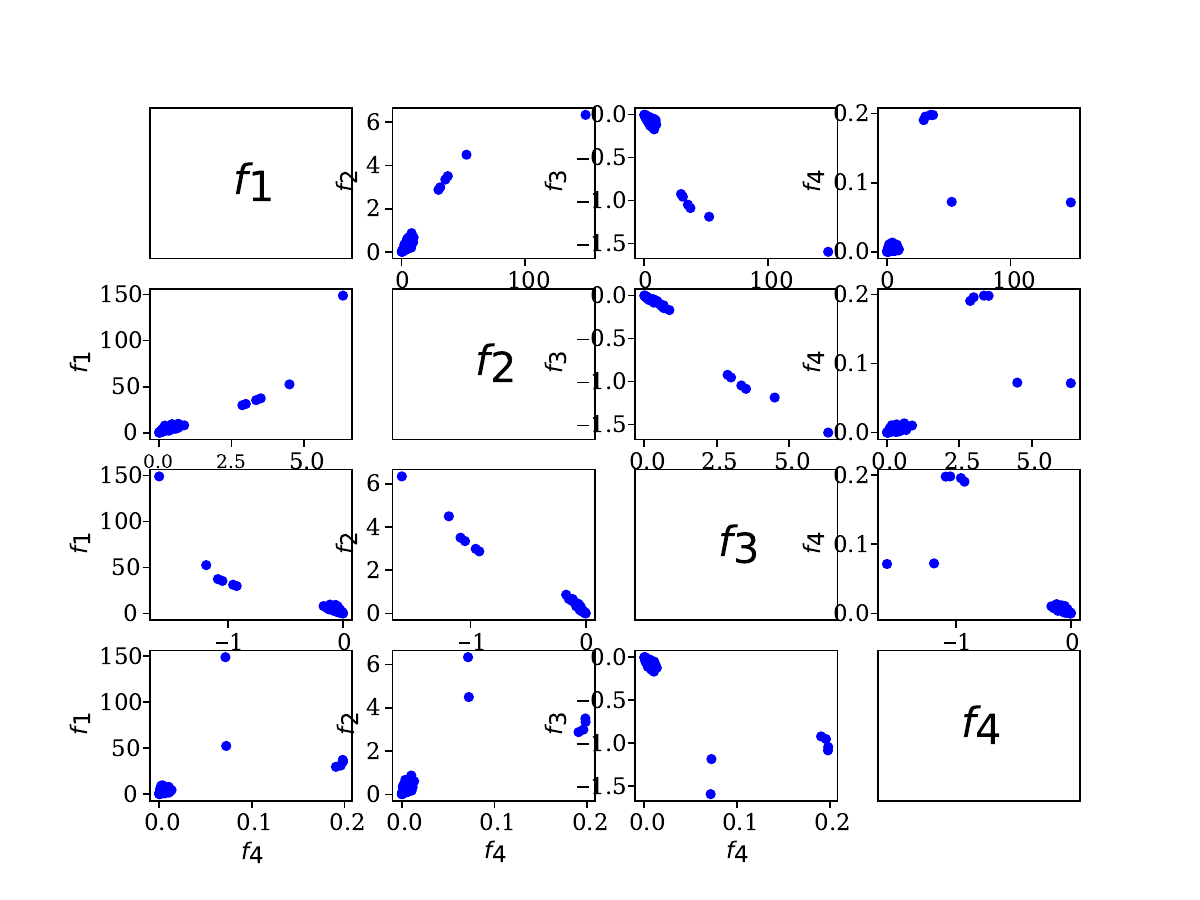}
        \caption{PSMOA}
        \label{fig:figure3}
    \end{subfigure}
    \caption{Performance comparison of NSGA-II, NSGA-III, and PSMOA at population size 100.}
    \label{fig:comp}
    \vspace{-1em}
\end{figure*}

We %simulate the PSMOA algorithm and 
compare PSMOA's performance with other well-known multi-objective optimization algorithms such as Non-Dominated Sorting Genetic Algorithm II (NSGA-II) \cite{996017} and NSGA-III. The simulation considers four objectives: replication cost $f1$, replication time $f2$, popularity $f3$, and load balancing $f4$. The parameters of the simulation include the number of nodes, data object sizes, storage capacities, network bandwidths, and user request patterns, which define the characteristics of the decentralized system.
%While PSMOA extends NSGA-III with policy support capabilities, the comparison with NSGA-II and NSGA-III serves several important purposes. First, these algorithms represent the state-of-the-art in multi-objective optimization and provide well-established performance benchmarks. Second, the comparison demonstrates how domain-specific adaptations (in this case, for data replication) can effectively guide the search process toward more relevant solutions while maintaining the robust optimization capabilities of the underlying algorithm.
Comparing PSMOA (which extends NSGA-III) with NSGA-II and NSGA-III demonstrates how our domain-specific policy adaptations  (in this case, for data replication) improve performance while maintaining the underlying optimization capabilities of these state-of-the-art algorithms.

Fig. \ref{fig:comp} presents a comparative analysis of NSGA-II, NSGA-III, and PSMOA in the objective space. The figure contains scatter plots of solutions for each pair of objectives which allows a comprehensive examination of the trade-offs between different objectives.
NSGA-II solutions are widely spread across the objective space which indicates a high level of diversity in the generated solutions. The scattered distribution may make it challenging for decision-makers to identify a clear set of optimal solutions. NSGA-III solutions are more concentrated in specific regions of the objective space, suggesting a focus on certain trade-offs between objectives. While this concentration may limit the exploration of the entire objective space, it can be beneficial when decision-makers have specific preferences or constraints.

\subsection{Performance Metrics}

To evaluate PSMOA's performance, we employ three widely used metrics. The Hypervolume (HV) measures the volume of the objective space dominated by the Pareto front and bounded by a reference point. A higher HV indicates better convergence and diversity of solutions. Given solutions $X$ and reference point $r$, HV is calculated as $HV(X,r) = volume(\bigcup_{x \in X} [f_1(x),r_1] \times ... \times [f_m(x),r_m])$
, where $f_i(x)$ is the $i$-th objective value.

The Generational Distance (GD) measures the average distance between the obtained solutions and the true Pareto front. A smaller GD indicates better convergence. GD is calculated as $GD = \left(\frac{\sum_{i=1}^n d_i^p}{n}\right)^{\frac{1}{p}}$, where $d_i$ is the minimum distance from the $i$-th solution to the true Pareto front, $n$ is the number of solutions, and $p = 2$ in our case.

The Inverted Generational Distance (IGD) measures both convergence and diversity. A smaller IGD indicates better overall performance. IGD is calculated as $IGD = \left(\frac{\sum_{i=1}^{|P^*|} d_i^p}{|P^*|}\right)^{\frac{1}{p}}$
, where $d_i$ is the minimum distance from each point in the true Pareto front to the obtained solutions.

\begin{table}[ht]
    \centering
    \caption{Comparison of results across different algorithms}
    \begin{tabular}{l|l|l|l}
        \hline
        \textbf{Algorithm} & \textbf{Hypervolume} & \shortstack{\textbf{Generational} \\ \textbf{Distance}} & \shortstack{\textbf{Inverted Generational} \\ \textbf{Distance}} \\ \hline
        NSGA-II & 0.57 & 148.74 & 3.76 \\ \hline
        NSGA-III & 0.41 & 67.74 & 5.61 \\ \hline
        PSMOA & 0.50 & 20.29 &  0.78 \\ \hline
    \end{tabular}
    \label{tab:table}
\end{table}

In Table \ref{tab:table}, PSMOA demonstrates balanced performance with an HV of 0.50, GD of 20.29, and IGD of 0.78. While its HV is slightly lower than NSGA-II, PSMOA achieves the lowest GD and IGD values, indicating solutions closest to the true Pareto front with good distribution in the objective space.
The performance metrics show that PSMOA achieves better convergence (GD: 20.29) compared to both NSGA-II (148.74) and NSGA-III (67.74). This demonstrates that policy support effectively guides the search toward solutions that are both mathematically optimal and practically relevant for data replication scenarios. The better IGD values (0.78 vs 3.76 and 5.61) further indicate PSMOA maintains good solution diversity while focusing on policy-compliant regions of the search space. The simulation results demonstrate that PSMOA excels in multi-objective optimization, efficiently balancing exploration and exploitation in complex search spaces.

\subsection{Scalability Analysis and Performance Considerations}

To evaluate PSMOA's scalability, we first analyze its computational characteristics. The algorithm's complexity scales with system size:

\subsubsection{Computational Overhead}
For a deployment with n nodes, the major computational costs are:
(a) Weight adaptation computation: $O(n \log n)$ (b) Reference point generation: $O(n)$, and (c) Solution evaluation: $O(n^2)$.

Our empirical testing demonstrates the effective performance of up to 80 nodes. The 4-7\% improvement in load balancing addresses the critical challenge of uneven data distribution in production grids. The algorithm's adaptive weight adjustment reflects real access patterns in scientific computing infrastructures, with adaptation periods matching typical transfer windows for large datasets. 
%Our empirical testing demonstrates the algorithm's performance characteristics up to 80 nodes, with results aligning well with real-world deployment scenarios. For instance, the 4-7\% improvement in load balancing performance directly addresses a critical challenge faced by production data grids, where uneven data distribution frequently leads to hotspots. The algorithm's adaptive weight adjustment mechanism mirrors the actual access patterns observed in scientific computing infrastructures like WLCG, where data popularity fluctuates based on research cycles, and our observed 1-3 hour adaptation period aligns with typical data transfer windows in high-energy physics workflows where datasets range from tens of gigabytes to several terabytes. 
However, scaling to larger deployments, such as those encountered in the WLCG with thousands of nodes, presents additional challenges due to the quadratic ($O(n^2)$) complexity. Such scale would require enhanced optimization strategies, potentially including hierarchical approaches for solution space exploration, relaxed real-time constraints for non-critical updates, and distributed computation architectures for parallel solution evaluation.

\subsubsection{Experimental Results}

\begin{figure*}[ht]
    \centering
    \begin{subfigure}[b]{0.32\textwidth} % Full column width
        \centering
        \includegraphics[width=\textwidth]{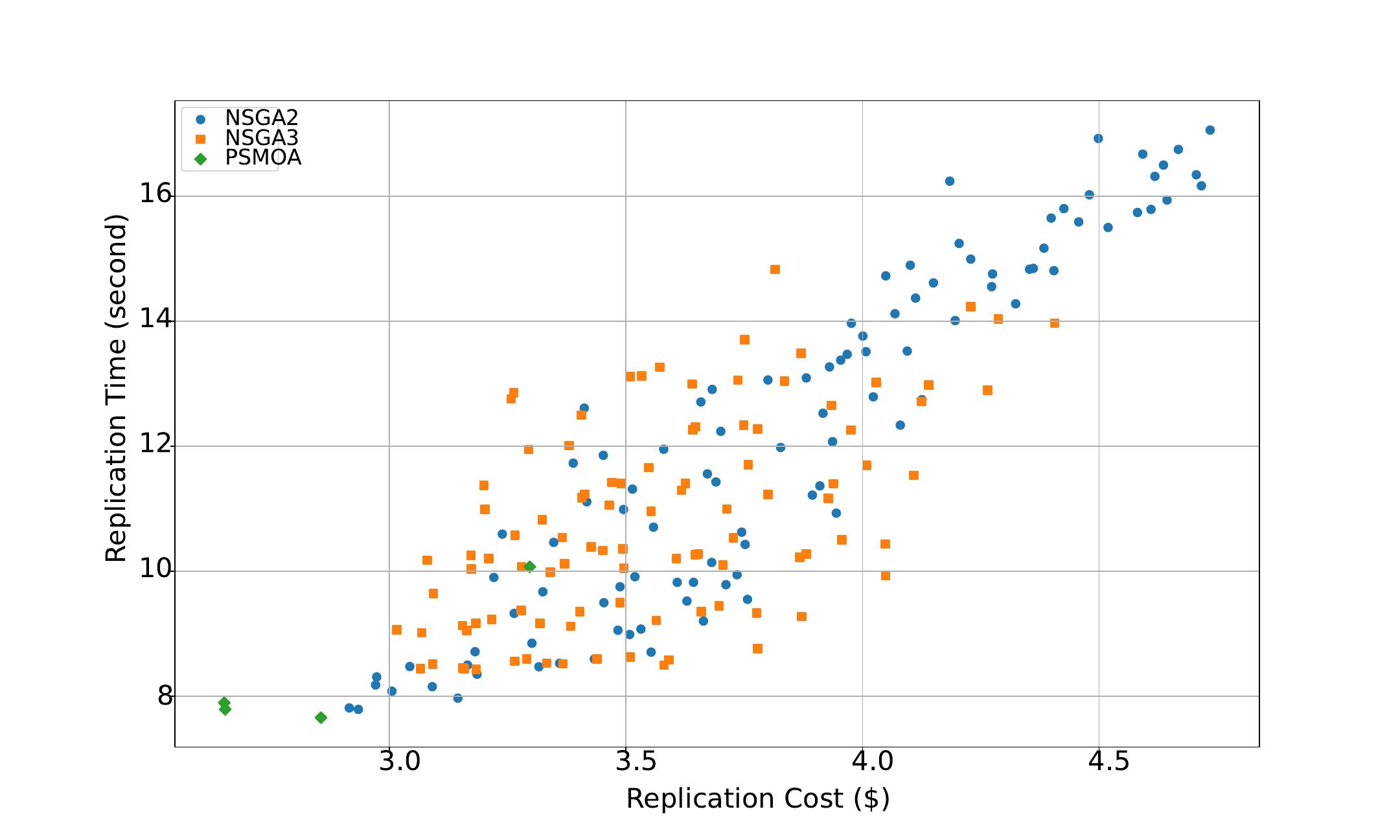}
        \caption{Small-scale: Cost vs. Time}
        \label{fig:s}
    \end{subfigure}
    \begin{subfigure}[b]{0.33\textwidth}
        \centering
        \includegraphics[width=\textwidth]{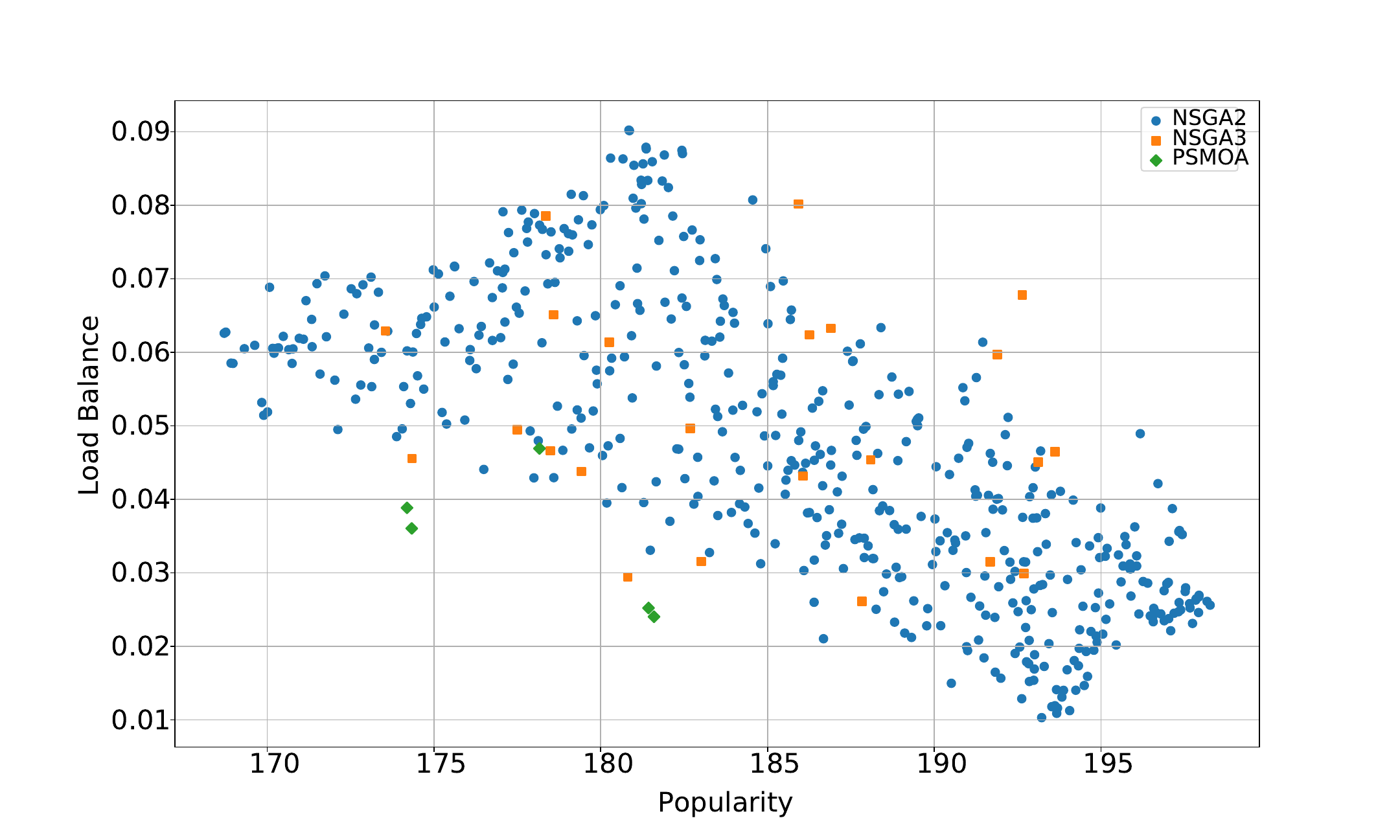}
        \caption{Medium-scale: Popularity vs. Load Balance}
        \label{fig:m}
    \end{subfigure}
    \begin{subfigure}[b]{0.32\textwidth}
        \centering
        \includegraphics[width=\textwidth]{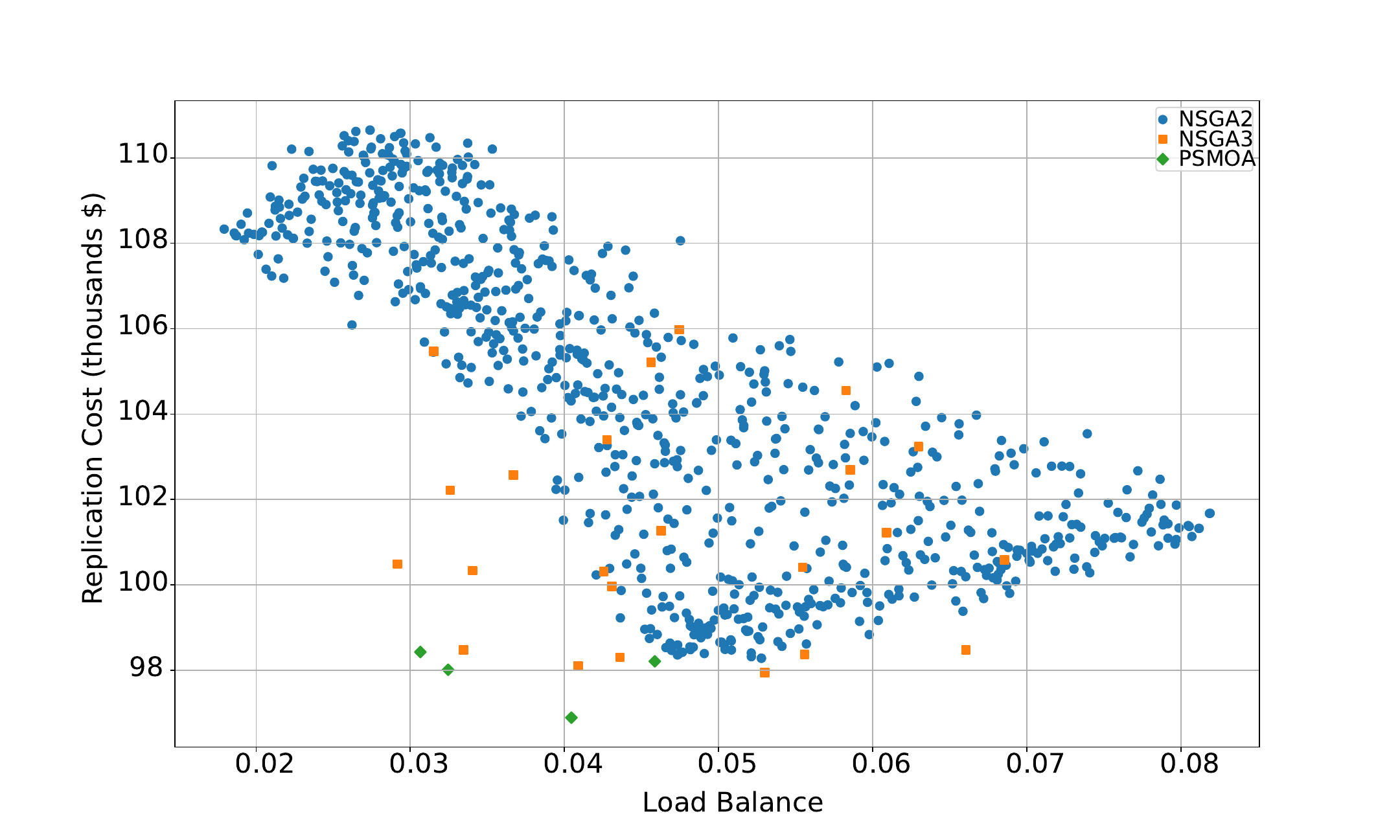}
        \caption{Large-scale: Load Balance vs. Cost}
        \label{fig:l}
    \end{subfigure}

    \caption{Pareto front comparisons across different system scales showing trade-offs between different objectives.}
    \label{fig:scale}
\end{figure*}

We conducted experiments with three system scales: small-scale (10 nodes, 20 data objects), medium-scale (30 nodes, 60 data objects), and large-scale (80 nodes, 150 data objects). Fig. \ref{fig:scale} illustrates performance comparisons across these scales using different objective combinations to highlight scale-specific challenges.

The Pareto fronts in Fig. \ref{fig:scale} demonstrate trade-offs in data replication across different scales. 
For small-scale deployments (10 nodes), PSMOA finds solutions in the optimal region, with replication costs from \$2.1 to \$3.4 and replication times between 7.5 and 10.0 seconds. These solutions clearly outperform both NSGA2 and NSGA3, which show more scattered solutions with higher costs (over \$4.5) and longer replication times (over 16 seconds). This concentrated distribution in the lower cost-time region demonstrates PSMOA's effectiveness in optimizing resource allocation even at smaller scales.

At medium scale (30 nodes), %the results reveal a complex interaction between popularity and load balance objectives. 
PSMOA manages popularity values between 174-183 while maintaining load balance variance between 0.024-0.047. 
This trade-off relationship reflects real-world access patterns where higher popularity (180-185) corresponds to increased load imbalance (0.04-0.06). 
Compared to NSGA2 and NSGA3's scattered solutions, PSMOA identifies focused solutions clusters that achieve higher popularity scores with better load balance. %This demonstrates PSMOA's effectiveness in finding solutions that prioritize both data access patterns and system stability at medium scale.

For large-scale deployments (80 nodes), PSMOA maintains tight control over replication costs (\$96.89-\$98.43) while effectively managing load balance variance (0.031-0.046) despite the exponential increase in solution space complexity, validating its capability to handle increased complexity while adhering to policy requirements, even at larger scales.

%A key challenge observed in scaling to larger deployments is the trade-off between optimization quality and computational overhead. While our current implementation shows promising results up to 80 nodes, further optimization would be required for federation-scale deployments like WLCG, which operates across thousands of nodes. This limitation suggests potential future work in developing hierarchical or distributed variants of PSMOA for extreme-scale deployments.

\subsection{Dynamic Policy Adaptation}
To evaluate PSMOA's dynamic adaptation capabilities, we designed a 24-hour experiment simulating real-world workload variations in a federated system. The experiment used 10 nodes with varying conditions throughout the day. We divided the daily cycle into three distinct periods: peak hours (8:00-17:00), off-peak hours (17:00-22:00), and night hours (22:00-8:00). During peak hours, we increased data access requests, averaging 200-300 requests per hour, along with stricter cost constraints reflecting budget policies. Off-peak hours featured moderate access patterns of 100-150 requests per hour with relaxed cost constraints, while night hours maintained minimal activity of 50-70 requests per hour.
\begin{figure}[h]

\centering
\includegraphics[width=\columnwidth]{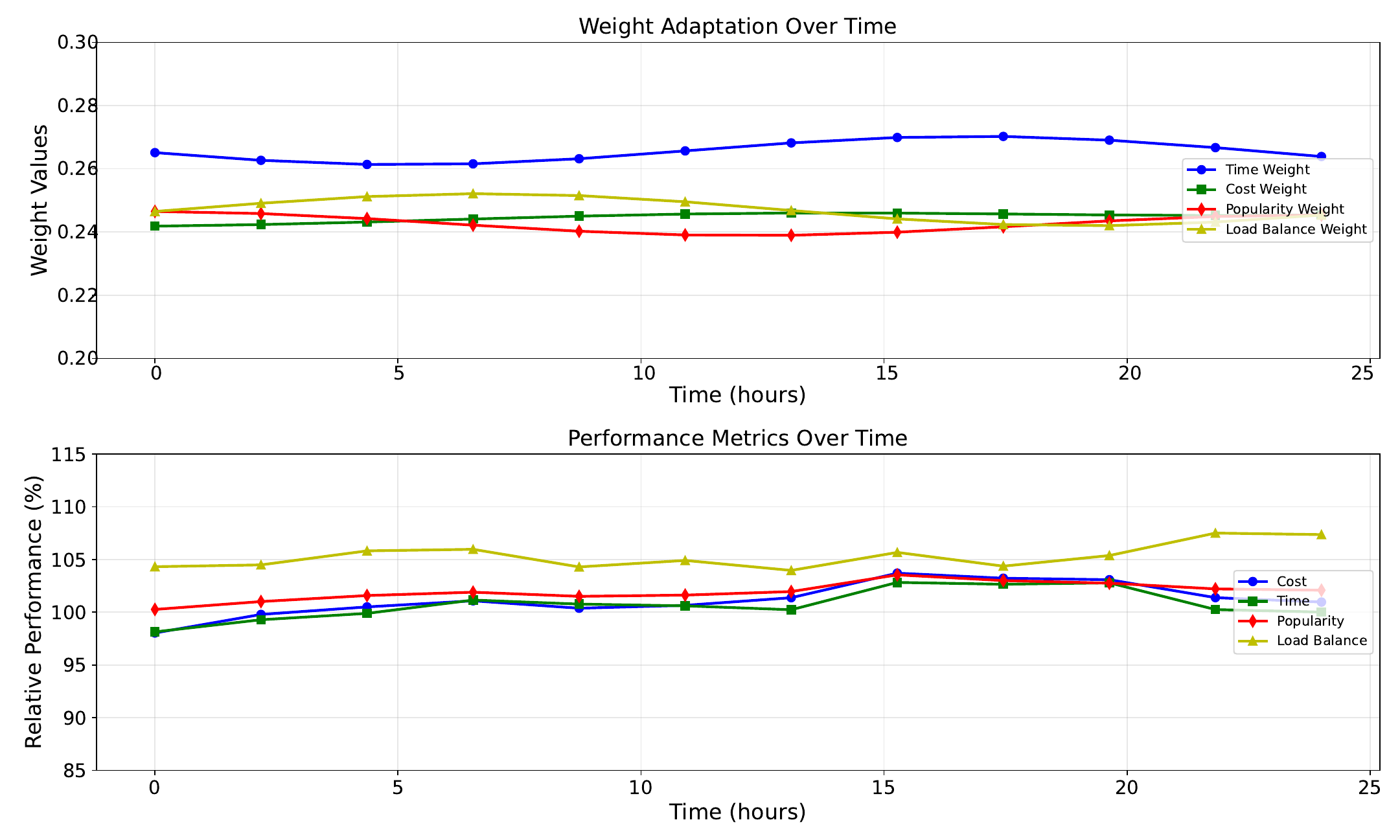}
\caption{Weight changes and performance over 24-hours}
\label{fig:simu1}

\end{figure}

Fig. \ref{fig:simu1} shows PSMOA's 24-hour dynamic policy adaptation through weight changes and resulting performance metrics. 
In the weight adaptation plot, we observe distinct patterns for each objective. The time weight shows the most dynamic behavior, varying between 0.26-0.27, with peaks during off-peak hours when network utilization is typically lower. Cost weight maintains a stable trajectory around 0.24-0.25, showing slight increases during the 10-15 hour period, reflecting business hours' cost consciousness. The popularity weight exhibits a gradual decline from 0.25 to 0.24 during the first 10 hours, followed by a recovery phase, indicating adaptation to changing access patterns. Load balance weight shows an inverse correlation with time weight, increasing during peak hours to maintain system stability.

The performance metrics plot uses 100\% as the baseline, representing initial system performance. Any value above 100\% indicated improvement over the baseline, while values below 100\% mean performance decline.
Using this reference point, the plot reveals how weight adjustments affect system behavior. Load balance performance consistently leads at 104-107\% of baseline (4-7\% improvement), showing gradual improvement, especially after hour 5, and demonstrating stability despite varying system conditions. The popularity metric maintains stable performance around 100-103\% (3\% improvement), exhibiting a slight positive correlation with its weight adjustments while showing resilience to weight fluctuations. Cost and time metrics remain well-controlled, staying within 98-103\% (3\% improvement) of baseline.

The relationship between weight adjustments and performance metrics reveals key insights about PSMOA's behavior. Weight changes typically precede performance improvements by 1-3 hours, demonstrating responsive yet stable adaptation to policy changes. This response time is critical in high-energy physics collaborations where data generation rates can exceed several GB/s, a 1-3 hour adaptation period allows the system to optimize replication strategies without disrupting ongoing data transfers. 
The algorithm maintains stability while continuously adjusting to policy requirements, as shown by consistent above-baseline metrics and balanced optimization across all objectives. This capability is very important in federated systems where organizations must balance multiple competing objectives while adhering to specific policy constraints.%, demonstrating PSMOA's ability to achieve load balancing improvements without compromising other operational metrics.

%% file: 6-Conclusions.tex
\section{Conclusions}

This work presents PSMOA, a policy-aware data replication algorithm that integrates organizational policies into multi-objective optimization. We demonstrate PSMOA's superior performance across both small- (10 nodes) and large-scale systems (80 nodes). 
%Despite these good results, several limitations need consideration in future research. 
Our current implementation assumes relatively static network conditions during optimization, which may not fully capture the dynamic nature of real-world federated environments. We also plan to investigate large-scale deployments with thousands of nodes where we might need to develop a %Additionally, a key challenge observed in scaling to larger deployments is the trade-off between optimization quality and computational overhead. 
%as system scale increases, the computational complexity of policy adaptation grows substantially, potentially affecting real-time decision-making capabilities in very large deployments. 
%While our implementation shows promising results up to 80 nodes, federation-scale deployments like WLCG, which operate across thousands of nodes, would require distributed optimization strategies to maintain performance. This suggests potential future work in developing 
hierarchical variants of PSMOA that can %effectively handle the increased complexity of extreme-scale deployments while preserving policy support capabilities.
decompose the global optimization problem into interconnected local optimizations, reducing computational complexity from $O(n^2)$ to $O(k \times (n/k)^2)$ where $k$ represents hierarchical levels while preserving policy support capabilities.